\documentclass[twocolumn, twocolappendix]{aastex631}

\usepackage[utf8]{inputenc}
\DeclareUnicodeCharacter{2212}{-}
\usepackage{graphicx}
\graphicspath{{./}{./CurrentFigures/}{./tex/figures/}}
\usepackage{amsmath}
\usepackage{afterpage}
\usepackage{enumitem}
\usepackage{xcolor}
\usepackage{float}
\usepackage{tabularx}
\usepackage{multirow}
\setenumerate{itemsep=0mm}

\usepackage{float}

\definecolor{newblue}{cmyk}{1,0.7,0,0}
\definecolor{newa}{cmyk}{2,0.3,1,0}

\usepackage{soul} 
\usepackage{amsmath}
\usepackage{amssymb}
\usepackage{xspace}
\usepackage{xifthen}
\usepackage{eso-pic}

\newcommand{\ruv}{R_{50, \rm UV}}
\newcommand{\ropt}{R_{50, \rm opt}}
\newcommand{\panoramic}{{PANORAMIC}\xspace}

\newcommand{\jwst}{\textit{JWST}\xspace}
\newcommand{\nircam}{{NIRCam}\xspace}

\newcommand{\sersic}{{S\'ersic}\xspace}

\newcommand{\numpy}{\code{NumPy}}
\newcommand{\scipy}{\code{SciPy}}
\newcommand{\matplotlib}{\code{matplotlib}}
\newcommand{\astropy}{\code{Astropy}}

\newcommand{\pysersic}{\code{pysersic}}

\newcommand{\stpsf}{\code{STPSF}}

\definecolor{forestgreen}{HTML}{228B22}
\definecolor{urlblue}{HTML}{000000}

\newcommand{\etal}{et al.\xspace}

\mathchardef\mhyphen="2D

\newlength{\dhatheight}

\newcommand{\code}[1]{\texttt{#1}\xspace}

\newcommand{\unit}[1]{\ensuremath{\mathrm{\,#1}}\xspace}

\newcommand{\pc}{\unit{pc}}

\newcommand{\e}{\unit{e^{-}}}

\newcommand{\secref}[1]{Section~\ref{sec:#1}}
\newcommand{\appref}[1]{Appendix~\ref{app:#1}}

\newcommand{\figref}[1]{Figure~\ref{fig:#1}}

\newcommand{\bandvar}[2][]{%
  \ifthenelse{\isempty{#1}}{\var{#2}}{\var{#2\_#1}}%
}

\newcommand{\var}[1]{\ensuremath{\texttt{\MakeUppercase{#1}}}\xspace}

\providecommand\physrep{\ref@jnl{Phys.~Rep.}}%
\providecommand\apjs{\ref@jnl{ApJS}}%
\providecommand{\jcap}{\ref@jnl{JCAP}}%

\defcitealias{Williams2024}{W25}

\begin{document}


\author[0000-0001-9978-2601]{Aidan P. Cloonan}\thanks{NSF Graduate Research Fellow}
\affiliation{Department of Astronomy, University of Massachusetts, Amherst, MA 01003, USA}

\author[0000-0001-7160-3632]{Katherine E. Whitaker}
\affiliation{Department of Astronomy, University of Massachusetts, Amherst, MA 01003, USA}
\affiliation{Cosmic Dawn Center (DAWN), Copenhagen, Denmark}

\author[0000-0003-0415-0121]{Sinclaire M. Manning}
\affiliation{Department of Astronomy, University of Massachusetts, Amherst, MA 01003, USA}

\author[0000-0003-2919-7495]{Christina C.\ Williams}
\affiliation{NSF National Optical-Infrared Astronomy Research Laboratory, 950 North Cherry Avenue, Tucson, AZ 85719, USA}
\affiliation{Steward Observatory, University of Arizona, 933 North Cherry Avenue, Tucson, AZ 85721, USA}

\author[0000-0002-5612-3427]{Jenny~E.~Greene}
\affiliation{Department of Astrophysical Sciences, Princeton University, 4 Ivy Lane, Princeton, NJ 08544, USA}

\author[0000-0001-5851-6649]{Pascal A. Oesch}
\affiliation{Department of Astronomy, University of Geneva, Chemin Pegasi 51, 1290 Versoix, Switzerland}
\affiliation{Cosmic Dawn Center (DAWN), Copenhagen, Denmark}
\affiliation{Niels Bohr Institute, University of Copenhagen, Jagtvej 128, Copenhagen, Denmark}

\author[0000-0001-8928-4465]{Andrea Weibel}
\affiliation{Department of Astronomy, University of Geneva, Chemin Pegasi 51, 1290 Versoix, Switzerland}

\author[0000-0003-2680-005X]{Gabriel Brammer}
\affiliation{Cosmic Dawn Center (DAWN), Copenhagen, Denmark}
\affiliation{Niels Bohr Institute, University of Copenhagen, Jagtvej 128, Copenhagen, Denmark}

\author[0000-0002-2380-9801]{Anna de Graaff}\thanks{Clay Fellow}
\affiliation{Center for Astrophysics, Harvard \& Smithsonian, 60 Garden St, Cambridge, MA 02138, USA}
\affiliation{Max-Planck-Institut f\"ur Astronomie, K\"onigstuhl 17, D-69117 Heidelberg, Germany}

\author[0000-0002-4684-9005]{Raphael E. Hviding}
\affiliation{Max-Planck-Institut f\"ur Astronomie, K\"onigstuhl 17, D-69117 Heidelberg, Germany}

\author[0000-0001-8460-1564]{Pratika Dayal}
\affiliation{Canadian Institute for Theoretical Astrophysics, 60 St. George St, University of Toronto, Toronto, ON M5S 3H8, Canada}
\affiliation{David A. Dunlap Department of Astronomy and Astrophysics, University of Toronto, 50 St. George St, Toronto ON M5S 3H4, Canada}
\affiliation{Department of Physics, 60 St. George St, University of Toronto, Toronto, ON M5S 3H8, Canada}

\author[0000-0002-8896-6496]{Christian Kragh Jespersen}
\affiliation{Department of Astrophysical Sciences, Princeton University, 4 Ivy Lane, Princeton, NJ 08544, USA}

\author[0000-0001-7673-2257]{Zhiyuan Ji}
\affiliation{Steward Observatory, University of Arizona, 933 North Cherry Avenue, Tucson, AZ 85721, USA}

\author[0000-0002-2057-5376]{Ivo Labbe}
\affiliation{Centre for Astrophysics and Supercomputing, Swinburne University of Technology, Melbourne, VIC 3122, Australia}

\author[0000-0003-1207-5344]{Mengyuan Xiao}
\affiliation{Department of Astronomy, University of Geneva, Chemin Pegasi 51, 1290 Versoix, Switzerland}

\author[0000-0001-6454-1699]{Yunchong Zhang} 
\affiliation{Department of Physics and Astronomy and PITT PACC, University of Pittsburgh, Pittsburgh, PA 15260, USA}

\submitjournal{\apjl}

\correspondingauthor{Aidan Cloonan}
\email{apcloonan@umass.edu}

\title{A \panoramic of UV-optical morphologies of ``Little Red Dots": Two groups of LRDs distinguished by UV half-light radius}

\shorttitle{\nircam Sizes of Little Red Dots in PANORAMIC}
\shortauthors{A. P. Cloonan \etal}

\begin{abstract}

Among the most remarkable results from \jwst is the discovery of abundant, compact, and very red sources in the early Universe known as ``Little Red Dots" (LRDs). The relative degree to which starlight and active galactic nuclei (AGN) drive the rest-frame UV and optical emission from LRDs remains unclear. With a large sample of LRDs selected photometrically from the pure-parallel PANORAMIC survey, we study their morphology as a function of rest-wavelength and find that the rest-UV light is typically more extended than the rest-optical. This result holds both when measuring LRD sizes with a single S\'ersic profile and when comparing the fraction of light from a point source via joint PSF+S\'ersic modeling. A shift occurs at the Balmer break, with LRDs becoming highly compact and unresolved ($R_{50,\rm{opt}}\lesssim100\;\rm{pc}$) in the rest-optical relative to the rest-UV. When splitting the sample at the Balmer break into those that are resolved and unresolved, a stacking analysis demonstrates that the latter are compact ($R_{50}\lesssim100\;\rm{pc}$) on average across the full rest-UV-optical spectrum. Conversely, those LRDs resolved at the break show extended UV emission ($R_{50,\rm{UV}}>200\;\rm{pc}$) on average. We find a similar dichotomy when repeating with a spectroscopic sample. Altogether, these results are consistent with the rest-UV emission driven by a combination of emission from starlight and a dense, dust-poor cloud of hydrogen gas enveloping an AGN. Differences between LRDs in the relative contribution from the AGN and starlight could reflect an ensemble of black hole seed masses, where a heavier seed produces an LRD of smaller $R_{50,\rm{UV}}$.

\end{abstract}


\section{Introduction}
\label{sec:intro}

Active galactic nuclei (AGN) are crucial tracers of galaxy evolution and supermassive black hole (SMBH, or BH henceforth) growth in a cosmological context \citep{Kormendy_2013}. In particular, feedback between AGN and their host galaxies may aid in regulating star formation as AGN inject heat into their surrounding gaseous environments \citep{Somerville_2008, Wellons_2023}. Little Red Dots (LRDs), a population of red AGN at high redshift discovered with \jwst in recent years, present a set of challenges to current models of galaxy and SMBH evolution. Not only are they very common \citep[e.g.,][]{Greene_2024, Kokorev_2024, Kocevski_2024, Matthee_2023}, but the mechanism(s) driving their emission are perplexing and difficult to explain \citep{Akins_2024, Setton2025}. Determining the physical nature of LRDs has important implications for how SMBHs form, grow, and interact with their environments in the early universe \citep[e.g.,][]{Matthee_2023, Kocevski_2024, Wang_2024, Greene2025}.

Several physical scenarios reproduce the observed LRD colors in the rest-UV and optical from \nircam photometric surveys, with differing contributions from starlight and AGN activity. These scenarios include: (1) a dust-reddened AGN with a small component of scattered UV light from the AGN \citep{Assef_2020, Pan_2021, Greene_2024, Labbe_2025}, (2) a red AGN with a faint component of extended, unobscured starlight from the host galaxy \citep{Barro_2023, Killi_2023, Matthee_2023}, and (3) a compact post-starburst galaxy, with a strong Balmer break produced by cooler, older stars \citep{Williams_2023}. To add to the confusion, LRDs are almost universally undetected in both X-ray with \textit{Chandra} \citep{Ananna2024, Kocevski_2024, Yue2024} and far-IR with ALMA and NOEMA \citep{Akins_2024, Setton2025, Casey2025, Xiao2025}. These non-detections point towards a smaller amount of hot coronal gas than would be expected for scenario (1). While scenario (3) is consistent with both a lack of X-ray and far-IR emission, it would also require exceptionally high star-formation efficiency and stellar mass density \citep[e.g.,][]{Baggen2024}. Another complication is that we do not find a spectral upturn in the rest-frame near-IR from hot dust associated with the dusty torus of an AGN \citep{Williams_2023, PerezGonzalez_2024, Wang_2024, Setton2025, Greene2025}, which presents a problem for scenarios (1) and (2).

More recently, another group of models has quickly grown in prominence, in which the central BH is buried within an optically thick, dust-poor medium at the center of a low-mass host galaxy \citep{Inayoshi2025, Kido2025, Liu2025, Liu2026, Naidu2025, Begelman2026}. In such cases, the extreme Balmer break and the UV-optical spectrum have a primarily non-stellar origin. The dense gaseous structure enveloping the AGN emits thermally with a photospheric temperature of $T\sim 5000\;\rm K$, and the weak UV and X-ray emission could result both from super-Eddington accretion \citep[e.g.,][]{Lambrides2024, Inayoshi2025, Liu2025} and from heavy gas absorption. The lack of dust is consistent with the lack of strong IR emission in LRDs. Assuming a BH-dominated continuum and Balmer break, the properties of the faint host galaxy such as stellar mass are therefore difficult to reliably determine \citep{Naidu2025}. Current dynamical measurements under this assumption typically prefer a host galaxy with a modest mass of $M_\star \sim 10^{9}\,M_\odot$ or even smaller \citep{Ji2025, DEugenio2025}.
Ultimately, observations indicate that the rest-UV-optical continuum of LRDs could be explained by a mixture of (buried) BH activity and starlight \citep{deGraaff2025, deGraaff2025b, Lin2025, Rusakov2025, Zhang2025}, but even if true, the strength of these processes relative to each other remain widely debated \citep{Ma2025}.

Morphology, especially in the rest-UV, may encode important clues for the underlying contribution from the LRD central engine and the host galaxy. A growing number of studies propose interpreting LRDs as a heterogeneous mix of BH systems and host galaxies, with both components varying in luminosity and other physical properties \citep[e.g.,][]{Barro2025, deGraaff2025b, Sun2026}. \citet{Golubchik2025} study a strongly lensed LRD and find a clear resolved component alongside the point source. Previous studies of LRD morphologies find on average that the rest-UV is more extended and/or asymmetric than the rest-optical \citep[e.g.,][]{Killi_2023, Baggen2025, Baggen2026, Chen2025, Jones2025, Rinaldi2025, Zhang2025, Ma2026}, indicative of some host galaxy emission. Morphological decomposition and half-light radius measurements of large LRD samples could therefore reveal the fractional contribution of each component to the UV-optical continuum.

In this paper, we conduct an analysis of multi-band morphology for a large sample of 181 LRDs with photometric redshifts ranging from $z_{\rm{phot}} \sim 4-9$, selected via \nircam photometry from PANORAMIC \citep[GO-2514, PIs: Williams, Oesch;][]{Williams2024}. 
We seek to quantify the difference between rest-UV and optical morphology with a statistically significant sample of LRDs and explore potential implications. We place statistical constraints on what drives the rest-UV and rest-optical continua without additional spectroscopic or multi-wavelength data, by relying on population statistics for the surface brightness profiles and sizes of LRDs from (observed-frame) 1.1$\mu$m to 4.4$\mu$m.

This paper is structured as follows. The \nircam imaging data and LRD sample selection are outlined in \secref{imaging_data}. We describe the morphology analysis in \secref{morphology} and results in \secref{results}, including size measurements and the characterization of LRD morphology as a function of rest-frame wavelength. We also explore the morphological properties of the LRD sample via stacking analyses in these sections, constraining how morphology changes with wavelength for the average LRD. In \secref{uv-balmer}, we investigate the LRD morphologies in the rest-UV and at the Balmer break in more detail, pairing the LRD sample from PANORAMIC with a smaller, spectroscopically confirmed sample of LRDs in order to reexamine our morphological results with this additional context. We explore the physical implications of the various tests and morphological models for the LRDs as a population of BH$+$galaxy systems in the early universe. Finally, we summarize our conclusions in \secref{summary}.

We adopt a cosmology of $h = 0.7$ and $\Omega_m = 0.3$, and all photometric magnitudes are given in the AB system.

\section{Imaging Data and Sample Selection}
\label{sec:imaging_data}

We search for LRDs in the \jwst Cycle 1 program PANORAMIC (GO-2514; PIs: Williams, Oesch; \citet{Williams2024}, henceforth \citetalias{Williams2024}), or ``Parallel wide-Area Nircam Observations to Reveal And Measure the Invisible Cosmos," a pure-parallel NIRCam imaging survey which covered roughly 530 arcmin$^2$. In this work we focus on the 430arcmin$^2$ with six or more broadband filters \citep[F115W, F150W, F200W, F277W, F356W, and F444W;][]{Rieke2023b}. In the pure-parallel mode, the observing setup is dependent on a primary program being executed simultaneously, meaning that information such as the location on the sky, the position angle of the camera, and the exposure time for an observation are predetermined. Consequentially, the imaging data from PANORAMIC consists of a disparate set of extragalactic pointings, at different locations and with different exposure times, though with a minimum of 42 minutes per pointing. The resulting $5\sigma$ observational depths in the F444W filter range from 27.8 ABmag at the most shallow to 29.4 ABmag at the deepest, with the full table of depths given in Table 3 of \citetalias{Williams2024}. Flux values in the source catalogs are measured through circular aperture-corrected photometry, with the image in each filter convolved to the point-spread function (PSF) in F444W prior to measurement. For fiducial measurements, an aperture radius of 0\farcs16 is adopted. 

The reduced data and science mosaics for the short-wavelength filters (F115W, F150W, F200W) are sampled at a pixel scale of 0\farcs02 per pixel, while those for the long-wavelength filters (F277W, F356W, F444W) are sampled at a pixel scale of 0\farcs04 per pixel.

\subsection{LRD Selection}
\label{sec:lrd_select}

\begin{figure*}
\centering
\includegraphics[width=15cm]{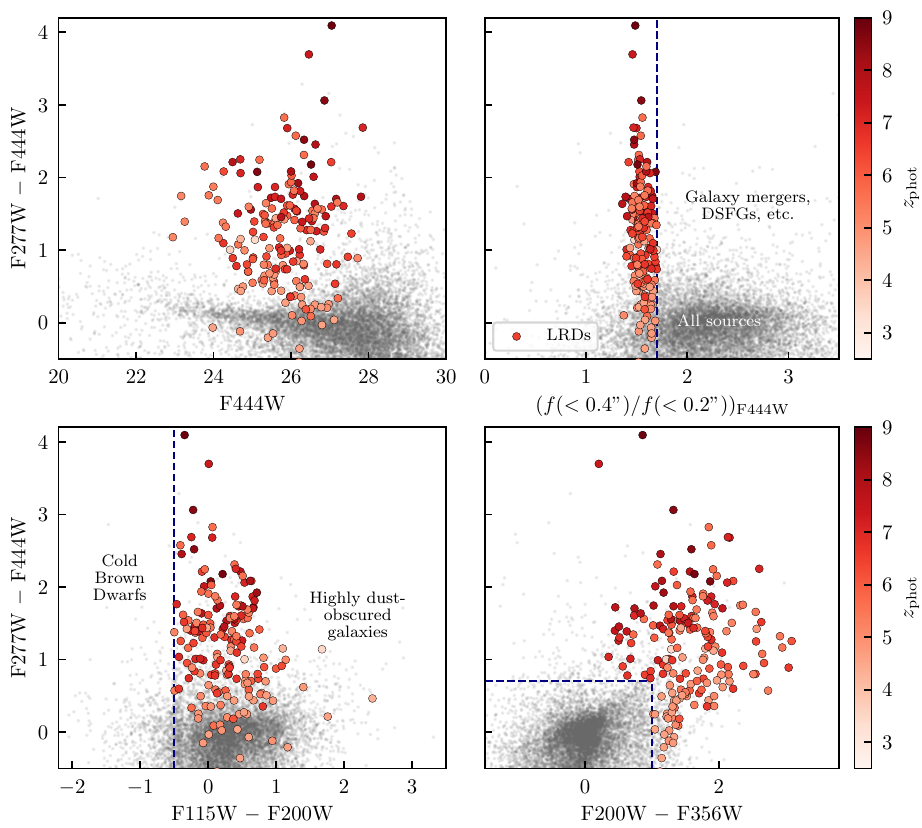}
\caption{We identify 181 reliable LRD candidates in PANORAMIC DR1 imaging via a selection process presented in \secref{lrd_select}, adapted from \citet{Kokorev_2024}. The most important criteria are shown here, with diagrams in color-magnitude \textit{(top left)}, color-compactness \textit{(top right)}, brown dwarf removal \textit{(bottom left)}, and $(\rm red1 \lor red2)$ \textit{(bottom right)}. The LRD sample spans a brightness range of $23 < \rm F444W < 28$. By selecting for red $\rm F277-F444W$ colors, we broadly select higher-redshift LRDs at $z>6$ (`red2'), whereas we find many missed LRDs at $z<6$ bluer in $\rm F277-F444W$ via red $\rm F200-F356W$ colors (`red1').}
\label{fig:sample-sel}
\end{figure*}

To ensure the reliability and accuracy of the science analyses on our LRD sample, we employ the selection method used by \citet{Kokorev_2024}, first proposed by \citet{Labbe_2025}. This method selects sources exhibiting ``v-shaped" SEDs at different redshifts, with one set of color cuts targeting sources at $z < 6$ (`red1') and another targeting sources at $z > 6$ (`red2'). The two sets are as follows:
\begin{align*}
	\rm{red1} =\ & \rm(F115W - F150W < 0.8) \\
	& \rm \land (F200W - F277W > 0.7) \\
    & \rm \land (F200W - F356W > 1.0),
\end{align*}
and
\begin{align*}
	\rm{red2} =\ & \rm( F150W - F200W < 0.8) \\
	& \rm \land (F277W - F356W > 0.6) \\
    & \rm \land (F277W - F444W > 0.7).
\end{align*}
Two additional cuts, one in compactness and one to remove cold brown dwarfs in the Milky Way, are added to this criterion.
\begin{align*}
	\rm compact =\ & f_{\rm{F444W}}(0\farcs 4) / f_{\rm{F444W}}(0\farcs 2) < 1.7 \\
	\rm{remove\ BDs} =\ & \rm( F115W - F200W > -0.5)	
\end{align*}
Putting these together, the final sample selection is:
\begin{align*}
	\rm LRD\ sample =\ & (\rm red1 \lor red2)  \ \land (\rm compact) \\
    & \land (\rm{remove\ BDs})
\end{align*}
From these criteria we find 183 candidates, out of which two are visually identified as components of merging galaxy groups and therefore removed.
The full resulting sample of 181 LRDs from PANORAMIC \citepalias{Williams2024} is shown in \figref{sample-sel} relative to the rest of the PANORAMIC galaxy catalog in terms of brightness (F444W), compactness, and several colors.

\jwst spectroscopic surveys demonstrate that such strategies requiring two optical colors, one UV color, and a compactness parameter reliably identify AGN. Building on the method's introduction by \citet{Labbe_2025}, \citet{Greene_2024} report a confirmation rate of $\gtrsim 80\%$ for the characteristic v-shaped spectrum and broad emission lines at $z \gtrsim 5$ in UNCOVER. Subsequently, \citet{Hviding2025} find a similar accuracy with a much larger spectroscopic sample in RUBIES. They also report $\sim$90\% accuracy and 50\% completeness for the selection method of \citet{Kokorev_2024}. The multiple red colors required ensure a steep red rest-optical continuum and remove emission line-boosted galaxies, which may appear red in one color but otherwise have bluer continua \citep{Endsley2023}. The compactness cut removes all varieties of extended sources. However, the analysis by \citet{Hviding2025} indicates that the method from \citet{Kokorev_2024} is more incomplete at lower rest-UV brightness than that from \citet{Kocevski_2024}, who fit power-law slopes in the rest-UV and optical to select LRDs. In doing so, they can find LRDs which are slightly fainter in the bluest \nircam filters more effectively. \citet{Hviding2025} note that both methods miss sources with the most extreme Balmer breaks such as the \textit{Cliff} \citep{deGraaff2025}. 

To summarize, the v-shape criterion reliably finds LRDs, but it is an insufficient description of the full diversity of intrinsic properties found in LRD SEDs. Our photometric sample is therefore incomplete and biased towards sources with substantial rest-UV flux.

\begin{figure}
\centering
\hspace{-1.1cm}
\includegraphics[width=9.5cm]{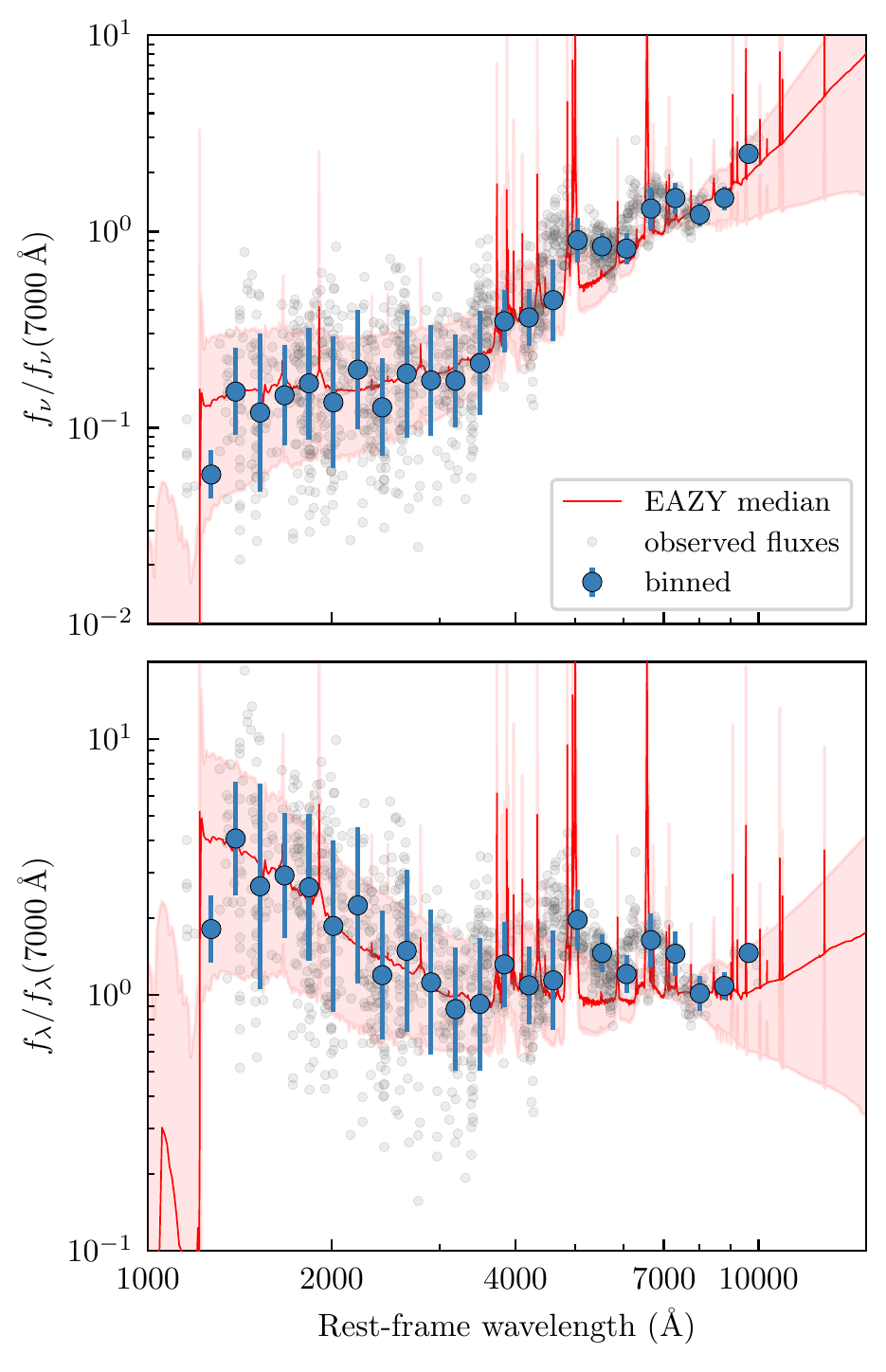}
\caption{With our sample selection for LRDs in PANORAMIC, we reliably find the expected `v-shaped' spectra seen in LRDs \citep{Kocevski_2024}, and we see spectral bumps in the survey data consistent with H$\beta$ and H$\alpha$. The median spectral energy distribution (SED) normalized at $7000\,\rm \AA$ with $\pm 1\sigma$ is shown in red, where we have fit individual LRDs with EAZY using a template based on the LRD from \citet{Killi_2023} to compute photometric redshifts. Individual LRD fluxes in each of the six broadband filters (converted into their respective rest-frames) are shown in grey, with a set of binned medians and logarithmic standard deviations shown in blue. We present the spectrum in units of both $f_\nu$ (\textit{top}) and $f_\lambda$ (\textit{bottom}), as the red optical color is shown more clearly in the former and the blue UV color in the latter.}
\label{fig:composite-sed}
\end{figure}

Given that the photometric criteria have been demonstrated to be accurate at selecting high-redshift galaxies, we re-compute each LRD's photometric redshift with EAZY \citep{Brammer2008}, imposing a prior of $3 < z < 10$ and using the \texttt{agn\_blue\_sfhz\_13} template\footnote{Summary of templates: \url{https://github.com/gbrammer/eazy-photoz/tree/master/templates/sfhz}}, which includes an AGN template based on the spectroscopically confirmed LRD reported by \cite{Killi_2023}. This way, we find each object's best-fit LRD redshift `solution' and also capture a more physical sense of the redshift uncertainty. 

We build a composite SED stack for the LRD sample by taking the median SED model flux at each rest-frame wavelength. We present a normalized version of the composite SED in \figref{composite-sed}, alongside the individual LRD catalog fluxes. By shifting to the rest-frame and normalizing to the model flux at $\lambda_{\rm rest} =7000\,\rm \AA$, we find evidence for broad emission lines and a large scatter in rest-wavelength among the LRD sample.

\section{Morphological Modeling}
\label{sec:morphology}

\subsection{LRD Sizes with Single \sersic Profiles}

To constrain the LRD morphology as a function of UV-optical wavelength, we model each selected LRD's surface brightness profile in all six broad-band filters, using the \pysersic code \citep{Pasha2023}. \pysersic is a newer Python package that is built to efficiently fit the light profile of galaxies with Bayesian inference, allowing for stronger constraints on posterior distributions and uncertainties in model parameters than other commonly used codes.

For each LRD, we fit a single \sersic profile \citep{Sersic_1963} to the light profile in each of the six broadband filters independently, using a Markov Chain Monte Carlo (MCMC). The background sky is fit as a flat constant simultaneously, where the constant should be approximately 0. The prior distribution is built using the following assumptions: a Gaussian prior for the flux with $\mu$ as the catalog's aperture-corrected flux value with $\sigma$ as the flux-error, a uniform prior in centroid allowing for up to 2 pixels in either direction about the center, and a uniform prior in half-light radius from 0.1 pixels to 20 pixels. The priors for ellipticity ($b/a$ between 0.1 and 1.0) and the S\'ersic index (between 0.65 and 8) are kept as the default in \pysersic. We then sample the Bayesian posterior distribution, from which we derive reliable uncertainties on LRD sizes.

When studying individual LRDs, we only consider a fit reliable if $S/N > 15$ in the photometric filter being modeled. \citet{Whalen2025} find that point sources can describe LRD light profiles for $S/N \lesssim 25$ regardless of intrinsic compactness, and recovering the half-light radius becomes significantly more difficult for $S/N \lesssim 15$. Note that the median S/N ratios for the LRD sample in each filter are as follows: 2.88 for F115W, 3.65 for F150W, 5.08 for F200W, 14.5 for F277W, 28.9 for F356W, and 31.7 for F444W. As such, only a small fraction of the total LRD sample will have reliable radius measurements in the rest-UV filters (i.e., F115W, F150W, F200W), whereas nearly all inferences will be robust in the rest-optical filters. We address this limitation via a stacking analysis.

\subsection{Point-Spread Functions}
\label{sec:psf_testing}

As described by \citetalias{Williams2024}, the current pipeline for the PANORAMIC data reduction simulates point-spread functions (PSFs) using the \stpsf \citep[formerly \code{webbpsf};][]{Perrin2014} tool in order to measure source fluxes and build the photometric catalogs. The motivation for using \stpsf as opposed to empirical PSFs extracted directly from the \nircam mosaics is that many pointings in the survey do not have a sufficient number of stars to enable a reliable PSF extraction. As such, \citetalias{Williams2024} opt to use \stpsf and rotate a given pointing's simulated PSF to match its position angle; they verify that the aperture photometry is robust to known systematics with PSF shapes from \stpsf \citep{Weaver2024}.

Conversely, inferred morphologies are quite sensitive to systematics associated with the PSF \citep{JWang2024}. LRDs are selected specifically to be compact, and therefore the structures of our PSFs are especially crucial to get accurate constraints on size and other morphological parameters. \citetalias{Williams2024} find that \stpsf is consistently narrower than the unsaturated stars in PANORAMIC, with an offset of $\sim 10\%$ for F444W within a $0\farcs16$ radius. However, as the PSF homogenization is only inaccurate for pixels within the minimum adopted aperture, such an assumption does not significantly impact the extracted photometry in principle (i.e., roughly the same total flux is conserved within the aperture, albeit with a different light distribution). Importantly, this is not the case when attempting to recover a robust light profile. As a consequence, using \stpsf for PSF convolution of the \sersic model would result in an overestimated size and a more extended profile \citep{Trujillo_2001, Liang2024}. 

To address this issue, we test the viability of using the empirical PSFs from the UNCOVER survey \citep{Weaver2024} on the PANORAMIC imaging by repeating the curve-of-growth analysis shown in Figure 6 of \citetalias{Williams2024}. In the F444W filter, at the fiducial aperture of $0\farcs16$, the UNCOVER PSF deviates from PANORAMIC stars by less than 1\% or 2\% with a median absolute deviation (MAD) of $<0.01\%$, and in F115W and F150W the deviation is several percent with a MAD of $\sim0.01\%$ (see \appref{psf-cog}). We find that the empirical PSFs from UNCOVER are sufficient for measuring the sizes of sources in PANORAMIC. Thus we use the UNCOVER empirical PSFs from \citet{Weaver2024} to measure LRD sizes from PANORAMIC.

\subsection{Mock Recovery Tests of the Resolution Limit}

The angular resolution limit for \nircam sources is a multivariate function that depends on observed wavelength, apparent brightness, and image depth. Different pointings in PANORAMIC vary widely in image depth, with F444W $5\sigma$ measurements ranging from 27.8 to 29.4 ABmag. To directly test the reliability of our size measurements and what counts as `unresolved' or `resolved,' we perform a simple test in which we fit a set of 1000 mock galaxies with known input sizes using the \pysersic pipeline, similar to \citet{Whalen2025}. With a set of true sizes and corresponding best-fit models, we can test for size recovery at different image depths and find the $S/N$ limit beyond which we recover radii reliably. Then, given sufficient $S/N$, we can find the value of true $R_{50}$ at which half-light radii become difficult to recover. We describe our method and results in \appref{res-limits}. 

\subsection{Joint PSF$+$\sersic Models}

After making a median stack of all LRDs in each of the six broadband filters (see \secref{stacking}), we fit each with a joint model of a \sersic profile plus a point source. For the priors, we allow the flux component to range from 0 to twice the catalog flux value, to allow for either of the two components to constitute most of the flux. We set a uniform prior on the \sersic half-light radius from 0.5 pixels to 10 pixels, and the other priors are the same as for the individual fits.

\section{Results for the Full Sample}
\label{sec:results}

\subsection{Wavelength Dependence of LRD Sizes}
\label{sec:wavedep}

\begin{figure}
\centering
\hspace{-1cm}
\includegraphics[width=9.25cm]{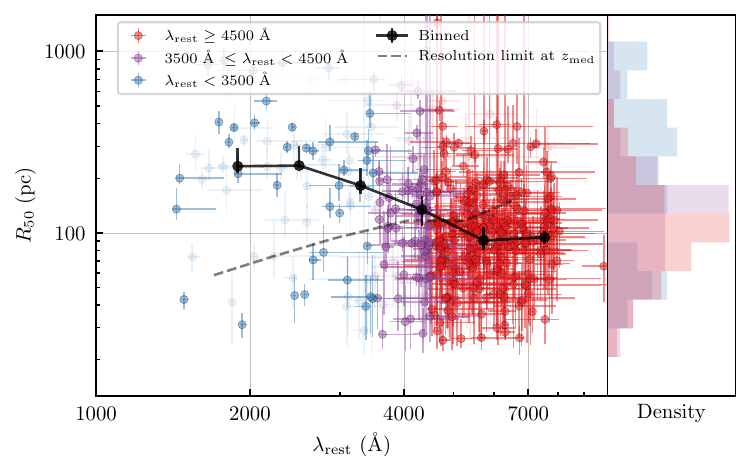}
\caption{The half-light radius for LRDs, with a small percentage of exceptions, decreases to very small radii at a rest-frame wavelength around the Balmer break, or just blueward of it. On average, as shown with the binned median points in black, the rest-UV flux is extended (with some scatter) while the rest-optical flux is not (with smaller scatter). For each filter, all radius measurements with a detection of $S/N >10$ are shown and sorted by rest-frame wavelength, with data points of $\rm 10 < S/N < 15$ being fainter and those of $S/N > 15$ being bolder. The resolution limit, shown as the grey dashed line, traces the upper bound in radius where median residuals for simulated sources satisfy $\left| \Delta\log R_{50} \right| > 0.02$ (see \appref{res-limits}), converted to kpc assuming the median redshift of $z\approx 5.79$. A normalized histogram for each of the three rest-frame wavelength bins is shown in the right-side panel.}
\label{fig:size-wavelengthLRDs}
\end{figure}

To investigate how the variation in LRD size across wavelengths, we present the half-light radii $R_{50}$ in units of pc as a function of the rest-frame wavelength in \figref{size-wavelengthLRDs}.  When plotting, we impose a S/N threshold of $S/N > 10$ for each filter separately, i.e., $R_{50}$ measurements with $S/N < 10$ are not shown. As such, far fewer $R_{50}$ measurements in the rest-UV are shown because the source is not sufficiently detected. Measurements with corresponding $S/N > 15$ are bolder as LRD morphological fits become more reliable at this detection level. We find that LRD sizes on average are larger at wavelengths bluer than the Balmer limit, i.e., at $\lambda \lesssim 4000\;\rm \AA$. Around the break, or just redward of the break, the LRD sizes consistently decrease to below the resolution limit that we computed in our simulations. However, this comparison is based on a smaller number of LRDs with brighter rest-UV flux, motivating a stacking analysis to account for this bias.

\subsection{Measuring the Average by Stacking LRDs}
\label{sec:stacking}

Due to low signal-to-noise and large uncertainties in photometric redshift, many individual size measurements have significant uncertainty, especially in the short-wavelength filters where LRDs are fainter. 
By stacking, we can get a better picture of a characteristic light profile than we can with the average of individual measurements of larger error.
To that end, we median-stack the LRD sample in each filter, i.e., in the observed frame. While it sacrifices information about individual sources, it also alleviates the issue of low S/N in the context of morphological fitting of LRDs \citep{Whalen2025}.

\begin{figure*}
\centering
\includegraphics[width=16cm]{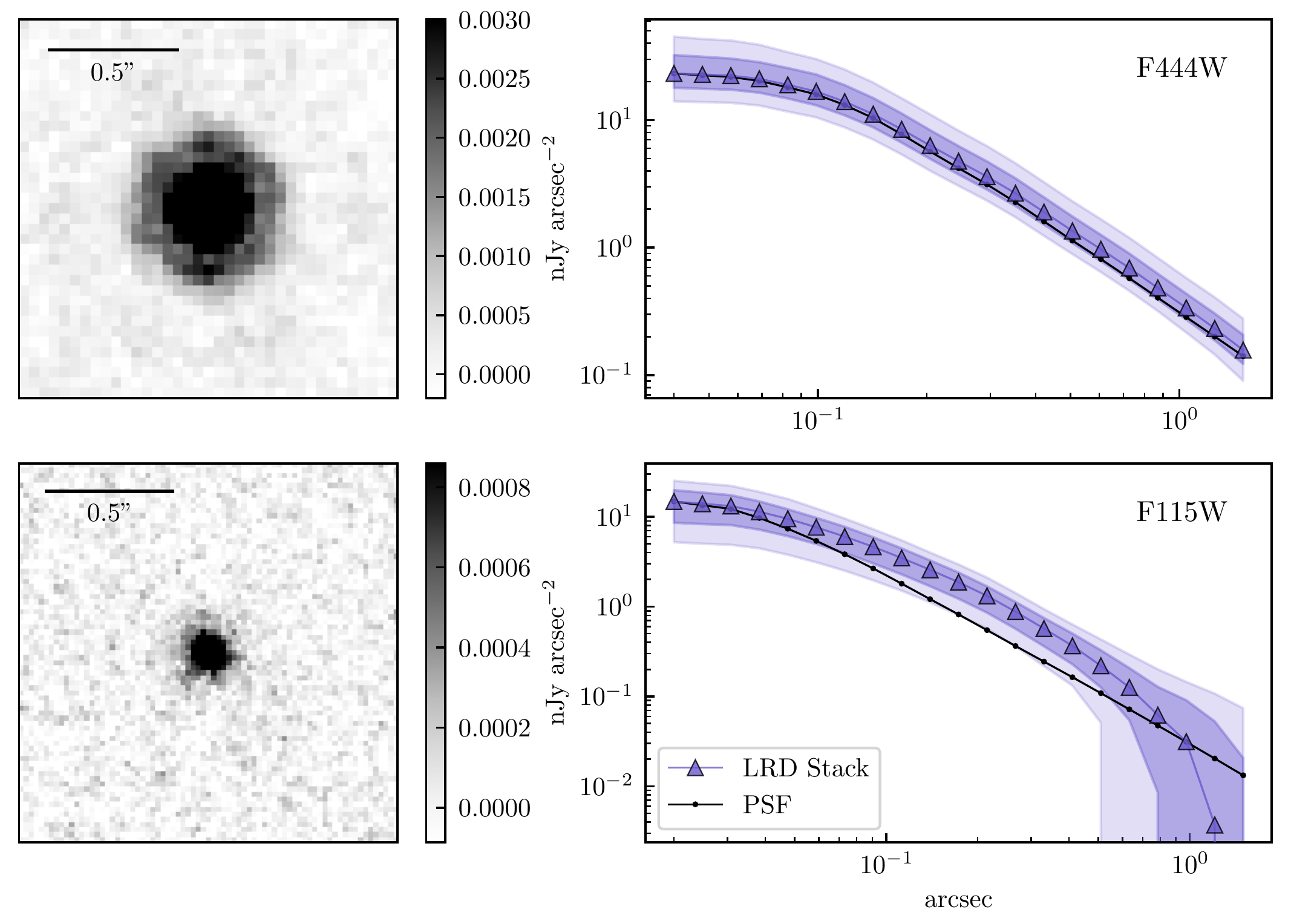}
\caption{Stacks of the full LRD sample in the `red' filters (e.g., F444W) are compact and their surface brightness profiles trace the PSFs closely. The `blue' stacks (e.g., F115W) are more diffuse and their surface brightness profile are shallower than the PSFs on average. We take $3\farcs \times 3\farcs$ cutouts of each LRD in each filter and then perform the stacking analyses. Zoom-ins of size $1\farcs5$ on the stacks for the F444W and F115W filters are shown on the left, and the surface brightness profiles are shown on the right. We perform bootstrapping when stacking to measure the distribution in each surface brightness profile, with the medians traced with light-blue, and the $1\sigma$ and $2\sigma$ uncertainties outlined in purple. To demonstrate whether LRDs on average are `point-like' or `extended' in a filter, we show the profile of the corresponding empirical PSF in black from the UNCOVER survey \citep{Weaver2024}.}
\label{fig:wstack-red-blue}
\end{figure*}

The stacking approach works as follows: a given LRD cutout is first re-centered by computing the flux centroid and shifting the cutout to be centered at the pixel containing the centroid. All other sources in each image are masked out. We exclude masked pixels from the stack, and we sigma-clip to remove $>3\sigma$ outliers to remove any remaining extended emission from foreground sources missed by the mask. Then, with an array of masked and recentered cutouts, we compute the final stack pixel by pixel, with each as the median of remaining, unmasked pixel values at its $(x,y)$ location. We show the median stacks of the full sample in the F444W and F115W filters and their surface brightness profiles compared to the corresponding filter PSFs in \figref{wstack-red-blue}. The stacks are in qualitative agreement with the trend observed from the individual size measurements, namely that LRDs are on average more extended and resolved in the bluer filters, while they become compact and unresolved in the redder filters (i.e., as steep as the PSF).

\begin{figure}
\centering
\hspace{-10pt}
\includegraphics[width=9.1cm]{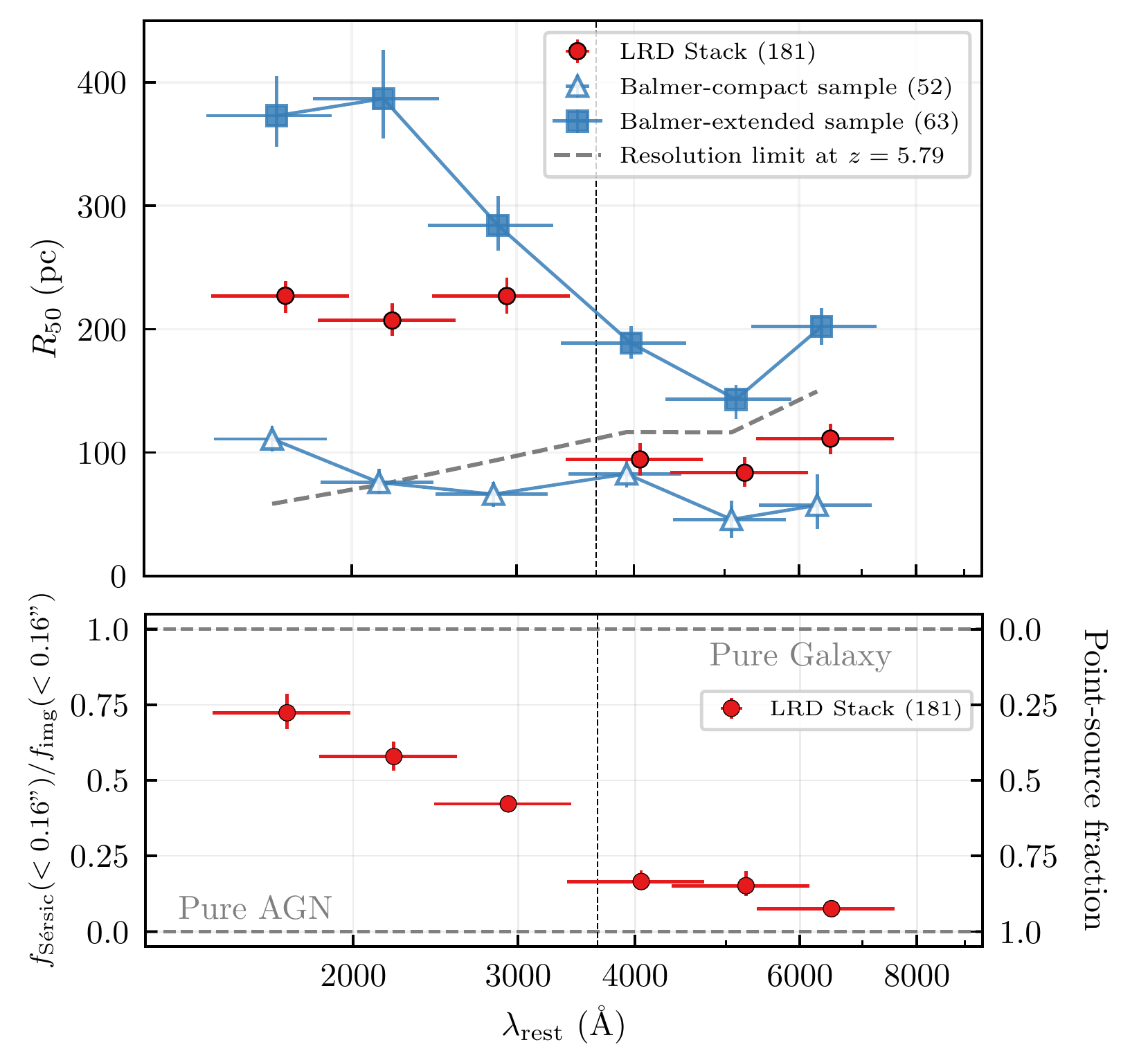}
\caption{Fitting a \sersic profile to each LRD stack (\textit{top}), we find that the characteristic half-light radii of LRDs in PANORAMIC sharply decreases with wavelength at the Balmer limit of 3645\,\AA. We find an analogous result by jointly fitting a point-source and a \sersic model to each LRD stack (\textit{bottom}), in that beyond the Balmer limit, the point-source accounts for most of the best-fit model aperture flux shown on the y-axis. In other words, an unresolved component becomes much brighter and dominates the total flux at wavelengths redward of the Balmer limit. From splitting the sample based on half-light radius at the Balmer break, we find two different morphological profiles shown in blue, with one following a similar transition at the Balmer limit but with increased size, and the other being compact across the rest-UV and rest-optical. These two profiles are consistent with two groups of LRDs, defined by the contribution of host galaxy light to the rest-UV. The x-axis errorbars are the $\pm 1\sigma$ wavelength values, measured from the redshift distribution, and are thus visual indicators of redshift variance in the stacked samples.}
\label{fig:stacking-sizes}
\end{figure}

The individual LRD sizes analyzed in \secref{wavedep} suggest that LRDs notably decrease in size at the Balmer break from 100s of parsecs in the rest-UV to $\lesssim 100$ parsecs in the optical on average. To further investigate and verify, we fit two morphological models to the stacked LRDs in each filter: one has a single \sersic component and the other is a joint PSF$+$\sersic profile. If the statistical trend in size with wavelength is real and this scenario is accurate, then we should find (1) that the half-light radii from the first model decrease dramatically at the Balmer break, and (2) that the point-source component in the second model becomes much more prominent at the break.

We show morphological fits to the stacked LRDs, assuming the two models, in \figref{stacking-sizes}. As anticipated, we find that the half-light radius suddenly drops off at the Balmer break and the stack becomes unresolved at longer wavelength, with a characteristic size of $R_{50} \approx 223^{+15}_{-14}\; \rm pc$ in F200W, blueward of the break. We find the same result for the joint PSF$+$\sersic models, where the fraction of flux attributable to the point source increases around the Balmer break. Numerically, the PSF fraction approaches 1 on average at $\rm \lambda_{rest} \geq 4000\,\rm \AA$ for the stacks, which are shown as red points in \figref{stacking-sizes}. 

\subsection{Interpretations}

We find that the LRD morphology depends strongly on wavelength. We observe more extended morphologies in the rest-UV whereas the vast majority ($\gtrsim 80\%$) of LRDs are unresolved in the rest-optical. (The latter is expected since LRDs are selected to be compact in F444W.) This transition from extended to point-like occurs around the Balmer break, consistent with the analysis carried out by \cite{Setton2024}, who demonstrate that the Balmer limit ($n=\infty \rightarrow n=2$) at $\rm 3645\;\AA$ is reliably the inflection point in the slope of LRD spectra. Our morphological results for the full sample indicate that extended sources\textemdash whether host galaxy starlight \citep[e.g.,][]{Killi_2023}, ionized gas \citep[e.g.,][]{Chen2025}, or both\textemdash contribute significantly to the rest-UV emission observed in LRDs. Then, the compact central source becomes prominent in the rest-optical, with a half light radius of $\ropt \lesssim 100 \pc$.

Complementary work by \citet{Jones2025} presents a morphological study of LRDs using high-resolution \nircam imaging to investigate their structural properties, namely black hole mass and stellar mass. Their analysis similarly attempts to disentangle compact nuclear emission from more extended host galaxy light across multiple wavelengths, and finds that while the rest-optical emission is frequently dominated by a compact unresolved component, extended emission is often present at shorter wavelengths. They further infer that when host galaxies are detected, they typically have stellar masses of $\sim10^{9}$–$10^{10},M_\odot$ and are relatively faint compared to the central source, contributing more to the total surface brightness in the UV. These results broadly support the picture emerging from our analysis that LRDs consist of a compact central engine embedded within a surrounding stellar system whose contribution becomes more clearly visible in the rest-UV.

We also find a similar basic picture when comparing our stacking analysis (\figref{stacking-sizes}) to that from \citet{Sun2026}, who measure the fraction of light from the central engine with spectroscopy. However, we do find a larger fraction ($\approx 50\%$ compared to $\approx 25\%$) of flux attributable to the point-source at $3000\;\rm \AA$. Furthermore, while we do observe the transition to a point source at the Balmer limit, we also find that the point-source fraction increases from $\sim 25\%$ to $\sim 50\%$ from $1500\;\rm \AA$ to $3000\;\rm \AA$. This change in the point-source fraction in the UV could reflect an intrinsically blue, unobscured host galaxy population.

Looking more closely at the morphology in the UV and near the Balmer break (i.e., $\approx 4000\rm;\AA$), we find that LRDs have a wider range of half-light radii at these wavelengths. In particular, about half of LRDs are unresolved at the break. This variance in half-light radii hints at a heterogeneity in the sample, as it may reflect a range of BH-to-galaxy light fractions. We explore this hypothesis further in the next section.

\section{Two LRD Populations Defined by the Balmer Break}
\label{sec:uv-balmer}

\subsection{Measuring UV and Balmer Break Half-Light Radii}
\label{sec:correlations}

We may expect the fraction of light from the central engine to depend on optical luminosity and other spectral observables \citep[e.g.,][]{Sun2026}. To look for more information about the possible host galaxy starlight or the central LRD engine, we inspect how the half-light radii of LRDs at the Balmer break ($\sim$4000\,\AA) correlate with the UV spectral slope $\beta_{\rm UV}$, optical color, and the optical luminosity $L_{5500}$ at 5500\,\AA. Using the photometric redshift, we identify the closest broadband filter to 2000 and 4000\,\AA. Given the redshift distribution for this LRD sample, the filters for 2000\,\AA\ sizes are either F115W, F150W, or F200W, and the filters for 4000\,\AA\ sizes are either F200W, F277W, or F356W. We only consider sources with $S/N > 10$ for the given filter, for which the uncertainty in individual size measurements is smaller.

We measure $L_{5500}$ and $\beta_{\rm UV}$ with the simple constraints on the LRD SEDs from EAZY (see \secref{lrd_select}). For $\beta_{\rm UV}$, we use the method outlined by \citet{Calzetti1994}. Briefly, we select 10 different specific ``fitting windows" in the range of $1200\,{\rm\AA} < \lambda < 2600\,{\rm\AA}$ in the UV spectrum \citep[see Table 2 in][]{Calzetti1994}, each of which avoids contamination from known intrinsic absorption features originating from stars and the interstellar medium. For each LRD, we take the rest-frame SED of the best-fit EAZY model and then fit a power-law slope to those 10 wavelength windows using a simple log-linear fit.

\begin{figure}
\centering
\hspace{-0.75cm}
\includegraphics[width=8cm]{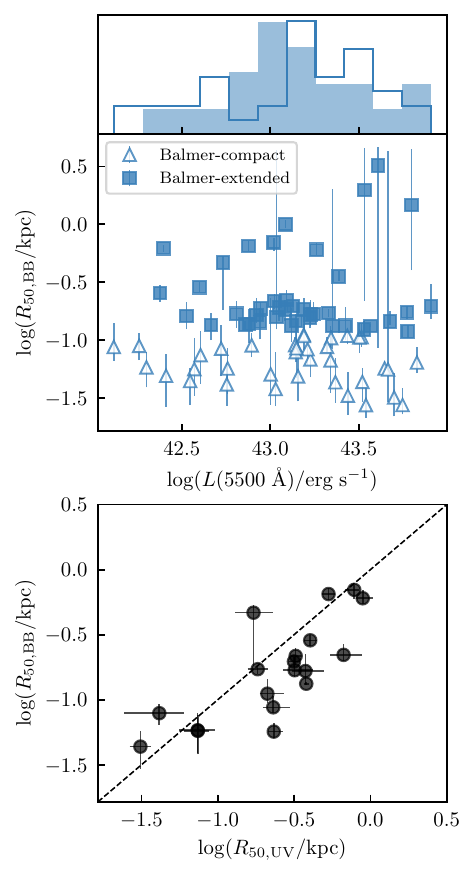}
\caption{We find no clear correlation between Balmer-break $R_{50}$ and optical luminosity $L_{5500}$ at 5500\,\AA (\textit{top panel}), but we do find a correlation between Balmer-break $R_{50, \rm BB}$ and rest-UV $R_{50, \rm UV}$ (\textit{bottom panel}). The luminosity distributions appear similar between the Balmer-compact and extended subsamples, though perhaps with a hint of bimodality in the Balmer-compact sample. When comparing $R_{50, \rm BB}$ and $R_{50, \rm UV}$, we show a smaller subset with sufficient $S/N$ in the rest-UV filter for a \pysersic fit.}
\label{fig:lum-sizes}
\end{figure}

In \figref{lum-sizes} we present both $L_{5500}$ and the UV half-light radius plotted against the Balmer-break half-light radius. We find no significant correlation of either LRD radius (i.e., when measured at the Balmer break or in the UV) with $\beta_{\rm UV}$, optical color, or $L_{5500}$. We do find a correlation between the UV radius and the Balmer break radius, indicating that LRDs which are more extended in the UV tend to be at the Balmer break as well. This correlation is based on smaller number statistics, however, as only $\sim10\%$ of the LRD sample has sufficiently high $S/N$ in the rest-UV filter. 

\subsection{Two LRD Stacks Split by Balmer Break Half-Light Radius}

Given the faint UV emission, we again perform a stacking analysis, this time to quantify any intrinsic correlation between half-light radius in the UV and at the Balmer break. We take LRDs from the sample with $S/N>10$ in the filter probing the break. Then, we split the resulting 115 LRDs into two based on whether $R_{50}$ is above or below the resolution limit in the Balmer-break filter. This gives two subsamples: 52 LRDs which are unresolved or compact at the break, and another 63 LRDs which are resolved or extended at the break. We then repeat the stacking analysis presented in \secref{stacking} to study each subsample's characteristic size as a function of wavelength.

When fitting a joint profile to the stacked LRDs unresolved at the break, we encounter a degeneracy in the rest-UV filters between the \sersic half-light radius and the PSF flux, reflecting a limitation highlighted by \citet{Whalen2025}. Their results indicate that the PSF fraction is difficult to measure if the `true' half-light radius of the \sersic component is less than 1 pixel, as the two components become increasingly similar in shape and difficult to separate. Therefore, we choose to fit a single \sersic profile to the two stacked subsamples, using the same priors as for individual LRDs (\secref{wavedep}). Fitting with single \sersic profiles ensures we can directly compare samples of potentially varying compactness.

We find that the two subsamples have significantly differing morphological profiles, with results presented as the blue symbols in \figref{stacking-sizes}. The LRDs which are unresolved at the Balmer break have much more compact rest-UV emission ($R_{50} \lesssim 100\; \rm pc$) when compared to the LRDs that are resolved at the break, for which we find $R_{50} \sim 300-500\; \rm pc$ in the rest-UV. Statistically, from one-tailed hypothesis tests resampling the posterior, their morphologies significantly differ at the $\sim 4\sigma$ level ($p \sim 10^{-5}$) in F115W and at $\sim 5\sigma$ ($p < 10^{-7}$) in each of F150W, F200W, and F277W. These results present evidence that these are two distinct groups of LRDs. Some LRDs are resolved at the Balmer break, and such sources emit extended UV emission on average. Other LRDs are compact and unresolved\textemdash or possibly barely resolved in F115W\textemdash across the UV-optical spectrum.

\subsection{Comparing with a Spectroscopic LRD Sample}
\label{sec:rubies}

One notable limitation of the described analysis with PANORAMIC is that the uncertainty in the photometric redshift distribution of the LRD sample may introduce a bias in the stacking analysis, if inaccurate. To see whether the size\textendash wavelength trend (\figref{size-wavelengthLRDs}) and the evidence for two LRD subsamples (\figref{stacking-sizes}) are significantly affected by any remaining redshift uncertainty,
we repeat both of these analyses using a spectroscopically confirmed sample of 36 LRDs from the RUBIES survey \citep{deGraaff2024, Hviding2025}. The range of optical luminosity probed by this sample is similar to PANORAMIC (see \secref{correlations}).

\begin{figure*}
\hspace{-0.75cm}
\includegraphics[width=19cm]{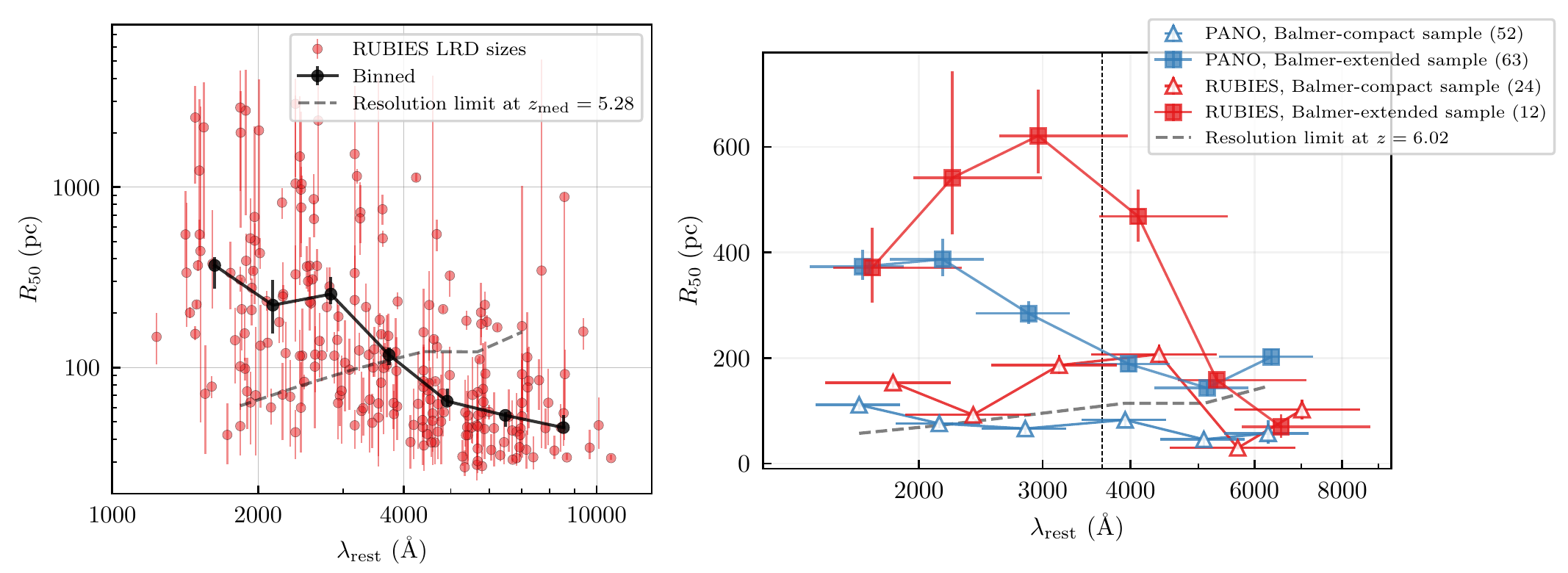}
\caption{Repeating morphological analyses with spectroscopic LRDs from RUBIES, we find a similar size-wavelength relation as with PANORAMIC and consistent results from stacking, though with slightly weaker statistical significance for the latter. \textbf{(Left)} The half-light radius in pc as a function of rest-frame wavelength for 36 LRDs from the RUBIES survey. As with PANORAMIC, there is significant scatter in the rest-UV sizes which are larger on average, while rest-optical sizes are mostly less than 100 pc and unresolved. \textbf{(Right)} The half-light radius in pc as a function of rest-frame wavelength for each Balmer break-compact and extended LRD stack, for LRDs from both PANORAMIC and RUBIES. For each survey, we find that the two subsamples differ in rest-UV size, albeit more dramatic for PANORAMIC given the tighter posterior distribution. The resolution limit is shown at the median redshift of 6.02 for the compact subsample in PANORAMIC.
}
\label{fig:size-rubies}
\end{figure*}

We find that the size-wavelength relation of individual LRDs from RUBIES is consistent with the result from PANORAMIC. The physical sizes of RUBIES LRDs are plotted against rest-frame wavelength in the left subplot of \figref{size-rubies}. Blueward of the Balmer break, the size distribution has significant scatter with an average of $\ruv \sim 300\;{\rm pc}$, while redward of the break the average falls below 100\;pc and the LRDs are largely unresolved. Intriguingly, just as with PANORAMIC, we find significant scatter in the half-light radius distribution at the Balmer break for this spectroscopic sample. While some of that scatter is likely due to low S/N in some cases \citep{Whalen2025}, if the scatter is intrinsic, that would indicate that this presence of both resolved and unresolved sources at the break is intrinsic\textemdash or, at least not a product of photometric redshift uncertainty.

We show the results of the stacking analyses for the two subsamples and the two surveys on the right of \figref{size-rubies}. For RUBIES, we find that the two subsamples differ from each other at $2\sigma$ or $>3\sigma$ significance depending on filter, a weaker but overall similar statistical trend as in PANORAMIC. LRDs unresolved at the break in RUBIES are again more compact in the rest-UV ($\ruv \lesssim 200\; \rm pc$) than their counterparts resolved at the break ($\ropt \sim 300-600\; \rm pc$). Two differences from the PANORAMIC results are that (1) each LRD subsample from RUBIES may have slightly larger sizes, and (2) the width of the $R_{50}$ posterior in the rest-UV filters for the subsample resolved at the break are larger. The statistical significance of the difference between compact and extended subsamples is therefore slightly weaker in RUBIES overall: we find a $\sim 3\sigma$ tension ($p \sim 10^{-3}$) in F115W, a $\sim 2\sigma$ tension ($p \sim 10^{-2}$) in F150W, and a stronger $\sim 5\sigma$ tension ($p \sim 10^{-7}$) in F200W. 

One potential source for the differences we find between PANORAMIC and RUBIES stacks is the difference in sample size. The RUBIES sample size is roughly 5 times smaller in number. If we randomly resample many sets of 12 sources from the subsample of 63 LRDs extended at the break in PANORAMIC, we find much broader posterior distributions in size when we concatenate together the different iterations' results. The statistical difference between the extended subsamples in RUBIES and PANORAMIC (i.e., comparing the red and blue squares in \figref{size-rubies}) decreases to $\sim 1\sigma$ in F200W and F277W. As such, we do not rule out sample size as a statistical driver behind RUBIES LRD stacks appearing more extended on average than PANORAMIC LRD stacks. Since the distributions in redshift and in optical luminosity $L_{5500}$ between the two surveys are similar, we do not expect either to account for the difference.

Despite these differences, the RUBIES results seem statistically consistent with the PANORAMIC results, and the takeaway is similar. We find tentative evidence of two morphological classes for LRDs as a function of wavelength, where one is more compact across UV-optical wavelengths than the other. 

\subsection{Interpreting Two LRD Subsamples}
\label{sec:discussion}

The spatial extent of Balmer break flux contains additional clues about the emission source(s) driving the observed properties of LRDs. From a stacking analysis of LRDs in PANORAMIC, we find that LRDs which are unresolved at the Balmer break are also compact in the rest-UV; these may be more BH-dominated and therefore unresolved across the rest-UV and optical. LRDs with extended flux at the Balmer break, on the other hand, are more extended and resolved in the rest-UV, indicating that the rest-UV and optical emission observed has a significant contribution from starlight in the host galaxy.

We find a statistically similar dichotomy in the LRDs from the RUBIES survey, though at slightly lower significance than for PANORAMIC due to larger model uncertainties for the subsample resolved at the break. While both samples are broadly consistent with the idea of two LRD subsamples, we note that RUBIES LRDs could be more extended on average than the PANORAMIC LRDs, at least in the rest-UV and at the Balmer break. Regardless of such differences, in both surveys we find that the UV half-light radius depends strongly on and how extended the LRD is at the Balmer break.

Putting everything together, we find that LRDs can be distinguished by the fraction of rest-UV and Balmer break flux attributed to an extended component, likely the LRD host galaxy \citep{Sun2026}. All LRDs are unresolved point-sources in the rest-optical, consistent with various frameworks for an optically thick gaseous envelope around a BH, described over the past year \citep{deGraaff2025, deGraaff2025b, Inayoshi2025, Naidu2025, Kido2025, Liu2025, Begelman2026}. Assuming each LRD is powered by a buried BH in a host galaxy, the fraction of UV light from the central engine will certainly depend on the accretion rate, the mass of the BH, the seed mass, and the host galaxy UV luminosity. For a UV-unresolved LRD, either the host galaxy is of low surface brightness, or the BH system is more massive and/or less optically thick along our line of sight. In both cases, the central engine accounts for nearly all of the UV-optical flux, and the total source is therefore compact and unresolved across the UV-optical spectrum. Conversely, a UV-resolved LRD may have a more luminous host galaxy, a less massive BH, or a denser central envelope. Here, the host galaxy contributes more significantly at bluer wavelengths since the LRD is faint in the rest-UV. LRD morphology in the UV and optical reflects these key properties, and it can place constraints on formation and early growth of SMBHs. For example, the presence of extended, UV-bright stellar companions could significantly affect BH growth in the central engine \citep{Baggen2026}.

\section{Summary}
\label{sec:summary}

The underlying physical processes driving the emission observed from LRDs are still not well understood, and the current hypothesized models \citep[e.g.,][]{Baggen2024, Inayoshi2025, Naidu2025} deserve further scrutiny. In this work, we present the analysis of a photometric sample of LRDs from the PANORAMIC survey's deep pure-parallel imaging \citepalias{Williams2024}, to study the morphological profiles of LRDs as a function of wavelength. In doing so, we build an empirical test for these physical scenarios. Our primary results and conclusions are as follows:
\begin{itemize}
	\item In the rest-optical, at wavelengths redder than the Balmer break ($\lambda_{\rm{rest}} \gtrsim 4000 \;\rm \AA$), the vast majority of LRDs are unresolved with physical sizes of $R_{50} \lesssim 100 \;{\rm pc}$. Meanwhile, in the rest-UV at bluer wavelengths than the break, LRDs are resolved and extended on average ($\approx 200-300\;\rm pc$, with a scatter of $\approx 100\;\rm pc$ in the distribution). We also find a similar pivot wavelength in the morphological profile as in the spectral shape \citep{Setton2024}. These results are consistent with a scenario where starlight from a host galaxy contributes to the rest-UV flux for a significant fraction of LRDs.
	\item A stacking analysis of the whole LRD sample is also consistent with rest-UV contribution from starlight. The LRD stack is highly compact and best represented by a point source in the rest-optical, while it is significantly more extended and better represented by a \sersic profile in the rest-UV.
    \item By separating the LRD sample based on whether or not they are resolved (i.e., extended) at the Balmer break, and then repeating the stacking analysis, we find that the subsample of LRDs unresolved at the break are significantly more compact in the rest-UV than the LRDs resolved at the break. The former are unresolved (or barely resolved) in all \nircam filters, consistent with a rest-UV and optical continuum dominated by a BH in a dense cocoon of gas. The latter are extended in the UV and at the break and unresolved in the optical, indicating significant contribution from starlight blended with the optically luminous point source. Interpreting the existence of these two subsamples, we favor the framework proposed by, e.g., \citet{deGraaff2025b} and \citet{Sun2026}, where LRDs are described by BHs enveloped in dense gas with a distribution of host galaxy light fractions.
\end{itemize}

We show in this work that morphology is a key tracer of the LRD emission source(s), and we demonstrate the power of pairing morphology at different wavelengths with spectroscopy to study LRD central engines and their environments. By combining both for larger samples, we can decompose flux from the central engine and host galaxy to provide new insight into the physical nature of LRDs. Deep and wide-area imaging is a key component of this future work, and we show that pure-parallel surveys provide the strong population statistics necessary to understand the diversity in the LRD population. Fully disentangling the structure(s) underlying LRDs will require understanding this diversity and resolving different components spatially as much as possible to complement the growing archive of LRD spectra.

\section*{Acknowledgments}

APC warmly acknowledges the support of the National Science Foundation through the NSF Graduate Research Fellowship Program.

This work is based in part on observations made with the NASA/ESA/CSA James Webb Space Telescope. The data were obtained from the Mikulski Archive for Space Telescopes at the Space Telescope Science Institute, which is operated by the Association of Universities for Research in Astronomy, Inc., under NASA contract NAS 5-03127 for JWST. These observations are associated with program JWST-GO-2514. APC, KEW, ZJ and CCW gratefully acknowledge support for program JWST-GO-2514 and JWST-GO-2561 provided by NASA through a grant from the Space Telescope Science Institute, which is operated by the Association of Universities for Research in Astronomy, Inc., under NASA contract NAS 5-03127. 

This work was supported by the International Space Science Institute (ISSI) in Bern, through ISSI International Team project \#25-659 `Little Red Dots, Big Open Questions'. The work of CCW is supported by NOIRLab, which is managed by the Association of Universities for Research in Astronomy (AURA) under a cooperative agreement with the National Science Foundation. AdG acknowledges support from a Clay Fellowship awarded by the Smithsonian Astrophysical Observatory. REH acknowledges support by the German Aerospace Center (DLR) and the Federal Ministry for Economic Affairs and Energy (BMWi) through program 50OR2403 `RUBIES'. PD warmly acknowledges support from an NSERC discovery grant (RGPIN-2025-06182).

\vspace{5mm}
\facilities{\jwst/\nircam}

\software{\numpy, \matplotlib, \scipy, \astropy, \pysersic, EAZY}

\bibliography{main}{}
\bibliographystyle{aasjournal}

\appendix

\section{PSF Curves of Growth}
\label{app:psf-cog}
\restartappendixnumbering

\begin{figure*}
    \centering
    \includegraphics[width=15cm]{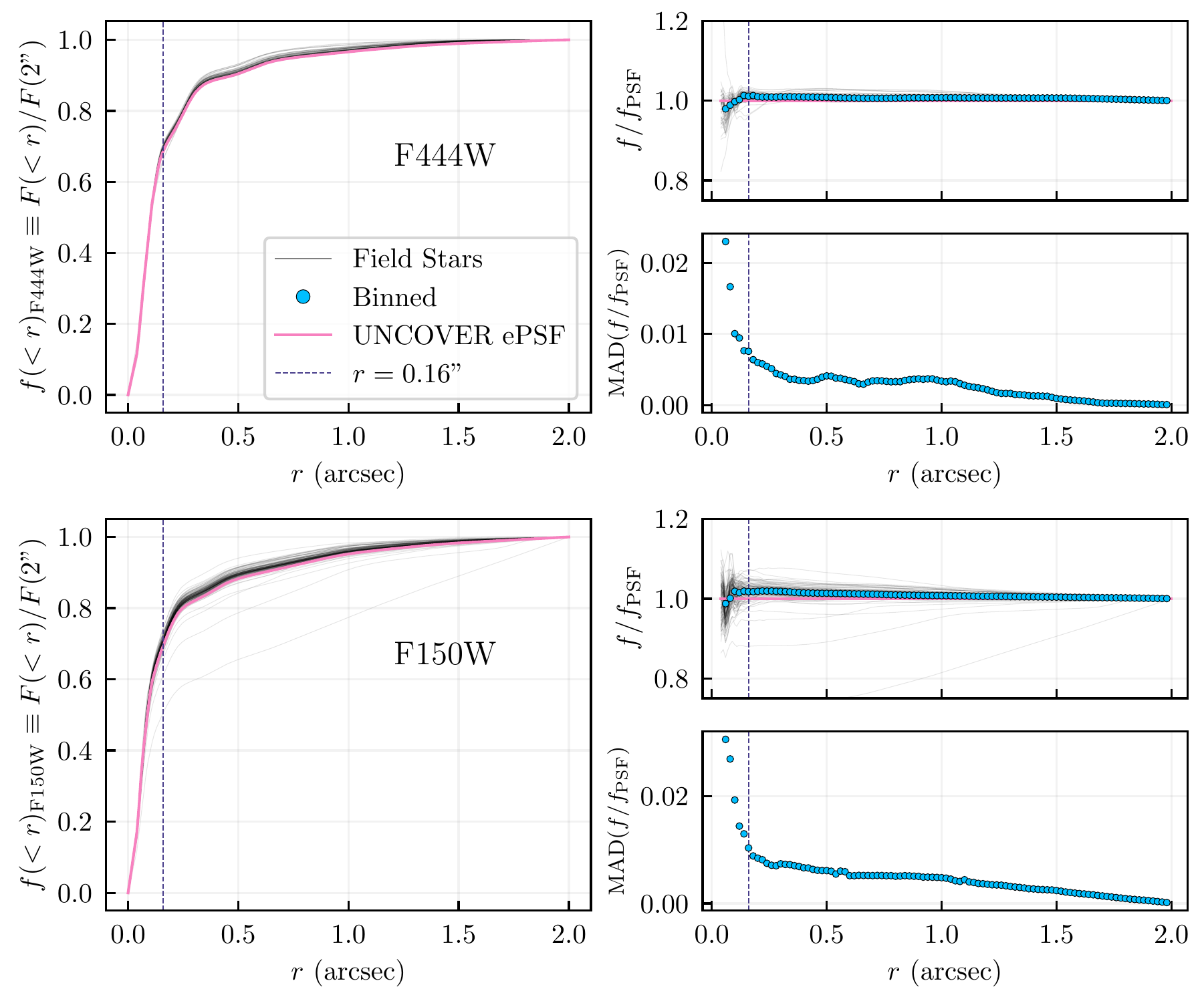}
    \caption{The curves of growth of UNCOVER PSFs provided by \citet{Weaver2024} are consistent with PANORAMIC stars to within a few percent of fractional error, with MAD of $\sim 1\%$ for a fiducial aperture with a radius of 0\farcs16. For each of F444W (\textit{top}) and F150W (\textit{bottom}), we show the aperture flux as a function of aperture radius, the fractional offset in flux $f$ of stars from the PSF as a function of aperture radius along with binned medians for stars (blue points), and MAD values for the binned medians.}
    \label{fig:psf-cog}
\end{figure*}

Here we present a comparison between the light profiles of empirical PSFs from UNCOVER \citep{Weaver2024} and of field stars in PANORAMIC, to verify the suitability of UNCOVER PSFs for our morphological analysis. The curve-of-growth results are visualized in \figref{psf-cog}. We find that convolution of a \sersic profile with the UNCOVER PSFs should not bias half-light radius measurements in PANORAMIC, whereas \citetalias{Williams2024} found that convolution with \stpsf PSFs would artificially increase inferred half-light radii.

\section{Resolution Limit as a Function of Wavelength}
\label{app:res-limits}
\restartappendixnumbering

Here we describe the method for estimating $R_{50}$ resolution limits in \nircam filters, which we use to distinguish resolved and unresolved flux in LRD light profiles. Then, we compare the results to other recently employed methods.

We first test for varying image depths. Each mock galaxy is randomly assigned an apparent half-light effective radius, and each has a \sersic index of $n=1$. The apparent radii are drawn from a log-uniform distribution between $-2.5 < \log R_{50} < 0$ in arcsec, and $S/N$ values are drawn uniformly in the range $5 < S/N < 100$. We find that varying the ellipticity of the \sersic profile does not affect recovery of the radius, and we also find similar recovery results for $n=4$. Then, for each mock galaxy we randomly select one cutout from our LRD sample and compute mock-background noise using the weight map of the cutout. The mock galaxy is then injected into the background noise realization. By using the weight maps from the cutouts of LRDs, we generate mock images with background noise akin to the \nircam images of LRDs in PANORAMIC. After fitting each mock image and comparing the true and recovered $R_{50}$ values, we identify the limit $(S/N)_{\rm lim}$ where residuals consistently deviate from 0 for $S/N < (S/N)_{\rm lim}$. In each filter we find that the accuracy and precision both decrease for $S/N \lesssim 15$, consistent with the conclusions from \citet{Whalen2025}. Since many LRDs are significantly fainter than this in bluer \nircam filters, we focus on population statistics rather than individual sizes when interpreting the model results on resolved and unresolved flux in the following sections.

\begin{figure*}
    \centering
    \includegraphics[width=15cm]{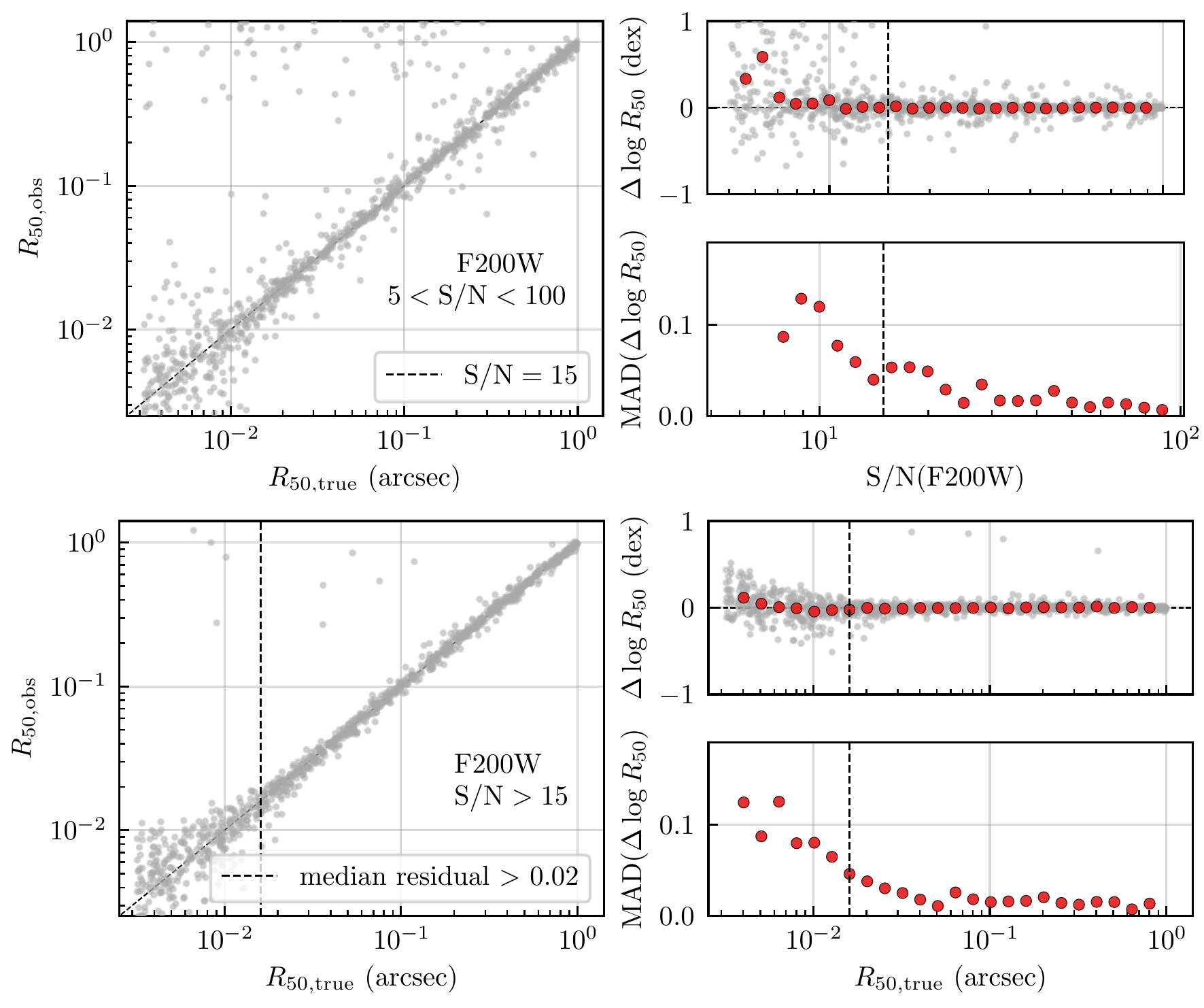}
    \caption{We find that half-light radii are accurately recovered for sources of $S/N>15$ (\textit{top half}), and we subsequently find a resolution limit of $\sim1/2$ to $3/4$ of a pixel for $S/N>15$ (\textit{bottom half}). Example simulation results are shown for F200W, and we find similar results for each filter. We show the true $R_{50}$ compared to the inferred $R_{50}$ (\textit{left}) and the residual $\Delta{\log R_{50}}$ between log-truth and log-inference as a function of $S/N$ (\textit{top half, top right}) and of the truth $R_{50}$ (\textit{bottom half, top right}). Binned median residuals are shown in red, highlighting the general trend, and we plot the median absolute deviation (MAD) of these binned $\Delta{\log R_{50}}$ points as a function of $S/N$ (\textit{top half, bottom right}) and of the truth $R_{50}$ (\textit{bottom half, bottom right}). The median residuals and MAD values each deviating from 0 for decreasing radii reflects a decrease in accuracy and precision, respectively.}
    \label{fig:sim-f200w}
\end{figure*}

Next, we estimate the resolution limit for our LRD light profiles as a function of observed wavelength, with results for F220W shown in \figref{sim-f200w} as an example. For each of the six bands, we repeat the simulation with all inputs unchanged except for $S/N$, for which we now draw uniformly from $15 < S/N < \rm(catalog\ maximum)$. We plot the residuals as a function of true $R_{50}$, and we define the resolution limit as the maximum radius at which the median residual satisfies $\Delta\log R_{50} > 0.02$. We choose this threshold as it corresponds roughly to the point at which the median absolute deviation of $\Delta\log R_{50}$ starts to increase, which implies that size measurements become less precise. For $S/N \gtrsim 15$, we find that the resolution limit steadily increases as a function of wavelength, from $\approx 0\farcs01$ in the F115W filter to $\approx 0\farcs025$ in the F444W filter, but it falls consistently in the range of 1/2 to 3/4 of a pixel.

We find a more conservative estimate of the resolution limit than \cite{deGraaff2025} who compare the LRD in that paper to a nearby star of the same apparent brightness, and we find a similar estimate to \cite{Hviding2025} through a different method. Instead of simulating mock galaxies, they fit S\'ersic profiles to stars and spectroscopically confirmed LRDs and adopt a brightness-dependent resolution limit based on the 95th percentile of the posterior distribution in size. While they find apparent radii of 0.25 pixels for many stars, they consistently find some scatter with some known point sources returning 95th percentile sizes of 0.5 pixels, or even as high as 1 pixel at the faint end.

\end{document}